\documentclass[a4paper,fleqn]{cas-dc}

\usepackage[numbers]{natbib}
\usepackage{tabularx}

\def\tsc#1{\csdef{#1}{\textsc{\lowercase{#1}}\xspace}}
\tsc{WGM}
\tsc{QE}
\tsc{EP}
\tsc{PMS}
\tsc{BEC}
\tsc{DE}

\begin{document}
\let\WriteBookmarks\relax
\def\floatpagepagefraction{1}
\def\textpagefraction{.001}

\shorttitle{Deep Learning-based Approach to Diabetic Retinopathy}

\shortauthors{H. Shakibania et~al.}

\title [mode = title]{Dual Branch Deep Learning Network for Detection and Stage Grading of Diabetic Retinopathy}

\author[1]{Hossein Shakibania}
\credit{Conceptualization, Methodology, Software, Investigation, Data Curation, Writing - Original Draft, Writing - Review \& Editing}

\affiliation[1]{organization={Department of Computer Engineering, Bu-Ali Sina University},
    city={Hamedan},
    country={Iran}}

\author[1,]{Sina Raoufi}
\credit{Conceptualization, Methodology, Software, Data Curation, Writing - Original Draft, Visualization}

\author[1]{Behnam Pourafkham}
\credit{Conceptualization, Methodology, Software, Investigation, Writing - Original Draft}

\author[1]{Hassan Khotanlou}
\credit{Writing - Review \& Editing, Supervision}

\author[1]{Muharram Mansoorizadeh}[orcid=0000-0002-7131-1047]
\credit{Writing - Review \& Editing, Supervision, Project administration}
\cormark[1]
\ead{mansoorm@basu.ac.ir}
\cortext[cor1]{Corresponding author}

\begin{abstract}
Diabetic retinopathy is a severe complication of diabetes that can lead to permanent blindness if not treated promptly. Early and accurate diagnosis of the disease is essential for successful treatment. This paper introduces a deep learning method for the detection and stage grading of diabetic retinopathy, using a single fundus retinal image. Our model utilizes transfer learning, employing two state-of-the-art pre-trained models as feature extractors and fine-tuning them on a new dataset. The proposed model is trained on a large multi-center dataset, including the APTOS 2019 dataset, obtained from publicly available sources. It achieves remarkable performance in diabetic retinopathy detection and stage classification on the APTOS 2019, outperforming the established literature. For binary classification, the proposed approach achieves an accuracy of 98.50\%, a sensitivity of 99.46\%, and a specificity of 97.51\%. In stage grading, it achieves a quadratic weighted kappa of 93.00\%, an accuracy of 89.60\%, a sensitivity of 89.60\%, and a specificity of 97.72\%. The proposed approach serves as a reliable screening and stage grading tool for diabetic retinopathy, offering significant potential to enhance clinical decision-making and patient care.
\end{abstract}

\begin{keywords}
Diabetic retinopathy\sep Fundus image analysis\sep Deep convolutional neural networks\sep Transfer learning
\end{keywords}

\maketitle

\section{Introduction}

Diabetic retinopathy (DR) is a severe eye condition that can profoundly impact vision and potentially lead to blindness if left untreated. DR is a complication of diabetes, a metabolic disorder that affects approximately 537 million people worldwide \cite{Federation2021}. According to the International Diabetes Federation (IDF), over one-third of individuals with diabetes experience DR, which is the leading cause of vision loss among working-age adults (20-65 years). The prevalence of DR is also increasing worldwide, with an estimated 103.12 million people affected in 2020, and it is expected to reach 160.50 million by 2045 \cite{Teo2021}.

\begin{figure}
    \centering
    \includegraphics[width=\linewidth]{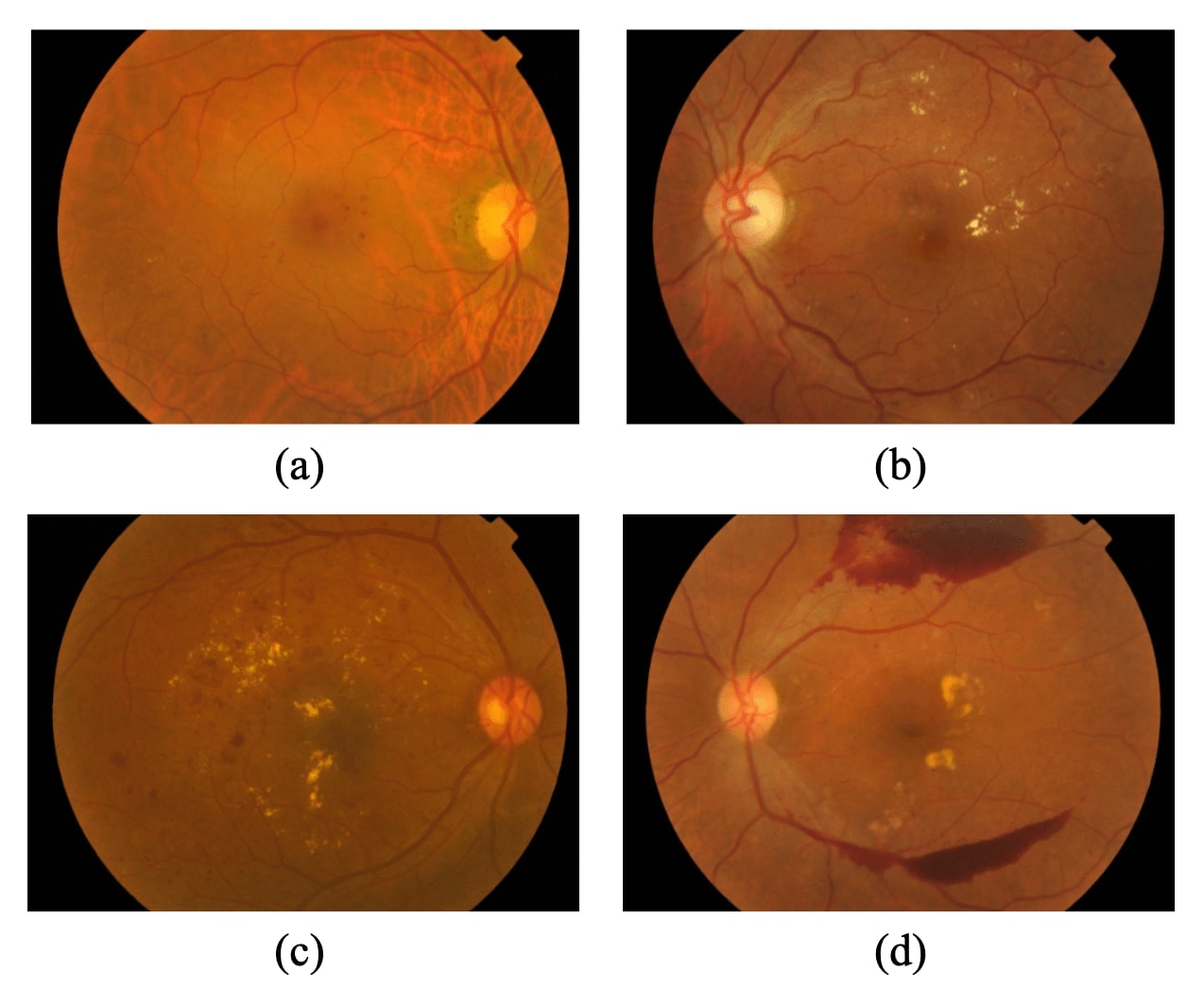}
    \caption{Four stages of DR: (a) Mild, (b) Moderate, (c) Severe, and (d) PDR obtained from APTOS 2019 dataset.}
    \label{fig:fig1}
\end{figure}

Generally, DR progresses through four stages:
\begin{itemize} \item \textit{Mild non-proliferative diabetic retinopathy (NPDR)}, the initial stage of DR referred to as background retinopathy, during which small bulges called microaneurysms develop in the retina's tiny blood vessels. These bulges can result in minor blood leakage into the retina. \item \textit{Moderate NPDR} is the second stage of the disease. At this stage, known as pre-proliferative retinopathy, the blood vessels in the eye become too swollen to nourish the retina properly. They may not carry blood as well as they used to, leading to physical alterations in the retina. \item \textit{Severe NPDR}, during which a considerable portion of the blood vessels within the retina experience blockage, resulting in a substantial decrease in blood circulation to the retinal tissues. In response, the retina may start sending signals to the body to develop new blood vessels. \item \textit{Proliferative diabetic retinopathy (PDR)} is the most advanced stage characterized by the growth of new blood vessels on the retina. However, these vessels are often weak and abnormal, leading to blood leakage into the eye, vision impairment, and potential blindness. Fig.~\ref{fig:fig1} provides a visual representation of the four stages of DR.
\end{itemize}

The necessity of early diagnosis and clinical intervention for DR cannot be overemphasized. A study by the national eye institute (NEI) found that timely treatment of DR reduced the risk of blindness by 95\% \cite{nei2019}. However, the problem is that DR can be asymptomatic in its early stages, making it difficult to diagnose. When symptoms such as blurred vision and floaters appear, the disease may have progressed to an advanced stage, making treatment less effective.

\begin{figure}
    \centering
    \includegraphics[width=\linewidth]{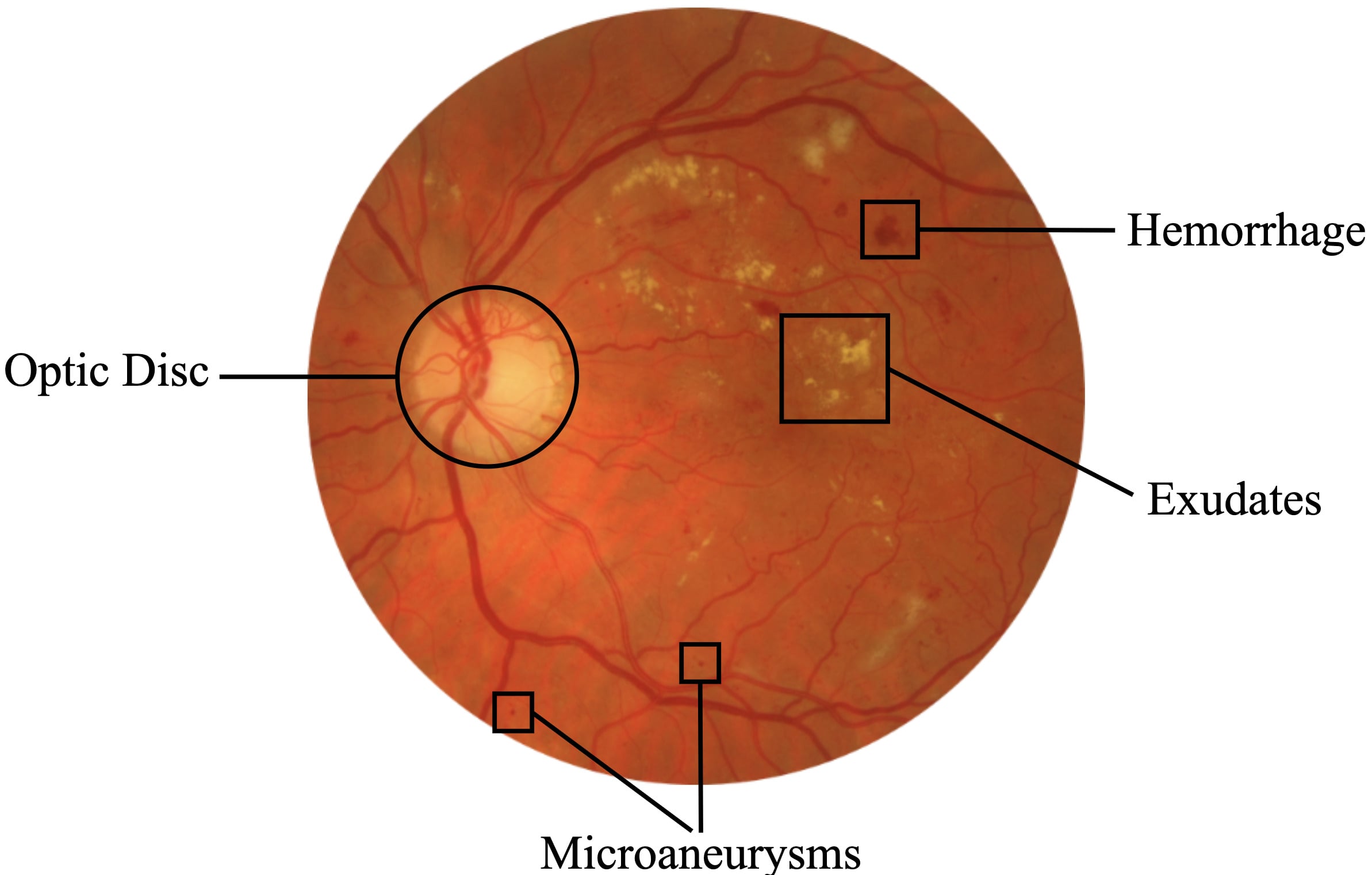}
    \caption{Fundus image with various lesions for DR classification.}
    \label{fig:fig2}
\end{figure}

Several approaches are used to diagnose DR, including dilated eye examination, optical coherence tomography (OCT), and fundus photography (FP). A dilated eye examination is the gold standard for detecting DR \cite{Rosses2017}. OCT is a non-invasive imaging method that can provide detailed and high-resolution images of the retina, allowing for the detection of changes in retinal thickness and morphology associated with DR \cite{Fujimoto2000}. OCT is particularly useful for detecting diabetic macular edema (DME), which refers to an increase in thickness of the macula, located at the center of the retina \cite{Sikorski2013}. However, OCT is better able to detect minor variations in retinal thickness and is less sensitive to retinal microvascular changes that occur in the early stages of DR \cite{Jiaa2015}. In contrast, FP can capture the microvascular changes associated with early-stage DR and has been used to assess and track the development of DR over a period of time \cite{Goh2016}. FP is particularly useful for detecting DR and grading its severity levels, which are divided into four stages based on the existence of different lesions like microaneurysms, hemorrhages, and exudates. These lesions are often visible in fundus photographs, as shown in Fig~\ref{fig:fig2}.

However, manual examination of these imaging modalities is time-consuming and prone to inter-observer variability. In addition, they are expensive, and not all healthcare facilities can access these technologies or the necessary expertise to diagnose DR effectively.

Another challenge in diagnosing DR is the lack of trained healthcare professionals in many parts of the world. According to the IDF, more than 50\% of people with diabetes are undiagnosed, and many of those diagnosed do not receive adequate eye care. This situation is particularly worse in low- and middle-income countries, with almost 90\% of undiagnosed diabetes, where resources for screening and treatment are limited \cite{Federation2021}.

Various initiatives have been launched to overcome these challenges to improve DR screening and treatment. One such initiative is telemedicine, where retinal images can be transmitted electronically to specialists for review, enabling remote diagnosis and treatment \cite{Vaziri2015, Boucher2008, Mansberger2013}. Another approach is using computer-aided diagnosis systems and artificial intelligence (AI) algorithms to analyze retinal images and detect signs of DR. AI-based systems have shown promising results in various fields of medical science. They can be used to detect such diseases accurately, quickly, and cost-effectively.

In recent years, machine learning (ML) algorithms have emerged as promising tools for the automated detection of DR. Various machine learning methods, including probabilistic models, linear and non-linear models, ensemble methods, and neural networks, have been applied to DR detection and grading with varying degrees of success \cite{Gulshan2016}. However, deep learning (DL) approaches, particularly convolutional neural networks (CNNs), have demonstrated superior performance in DR detection compared to classical machine learning algorithms \cite{Ting2017}.

Deep learning algorithms can automatically learn features from raw data without explicit feature engineering, making them well-suited for image-based applications such as DR detection \cite{LeCun2015}. CNNs have been shown to effectively learn hierarchical features from retinal images, leading to improved performance in DR detection \cite{Wan2018, Gargeya2017}. Furthermore, the ability of CNNs to manage substantial volumes of data and generalize to new data makes them an ideal tool for automating DR detection and stage grading.

Although numerous computer-aided methods have been employed to tackle the challenge of accurately detecting and grading DR, significant limitations persist, prompting a need for improved accuracy. Google's research indicates that even among retinal specialists, disagreements exist regarding the grading of DR from medical images {\cite{Krause2018}}. Therefore, we are motivated to propose a novel approach that builds upon prior works, responding to the critical need for advancements in diagnostic tools to confront the complexities associated with this debilitating condition. One predominant challenge in this task is the unbalanced nature of available datasets. Acknowledging this issue, we aimed to tackle the data challenge through innovative methodologies, something that most state-of-the-art methods have not attempted. We also anticipate that the utilization of two feature extraction branches, combined intelligently, will provide us with a robust representation of the intricate data patterns at each stage of the disease. Our motivation extends beyond technical advancements; we aim to use deep learning not only to enhance the accuracy of diagnosis and determine appropriate treatments for each stage of the disease but also to improve the overall patient experience and outcomes.

In this paper, we present our approach for detecting and grading DR, which leverages deep learning algorithms. Specifically, we utilize a transfer learning approach by using the pre-trained weights of two state-of-the-art deep learning models. We develop a dual-branch model that integrates features from multiple levels of abstraction for more accurate DR detection and stage grading. We develop a selectively merged and augmented dataset to train our model, strategically addressing the imbalances and data quality issues prevalent in existing datasets. Our experiments are mainly conducted on the APTOS 2019 \cite{kaggle2019}, a widely-used dataset for DR detection and grading, provided by the Asia Pacific Tele-Ophthalmology Society (APTOS) and hosted on the Kaggle platform for competition in 2019. Through extensive experiments, we prove the efficacy of our proposed approach and compare it with existing state-of-the-art methods in terms of several evaluation metrics. Our approach can potentially improve the early detection and severity grading of DR, leading to better clinical results.

\section{Literature review}

Diabetic retinopathy has attracted considerable interest among many researchers in the literature. Several studies have been conducted to develop automated approaches for diagnosing and grading the disease at different stages. Diabetic retinopathy classification is generally divided into two main types: binary classification and multi-class classification. Binary classification focuses on detecting the presence or absence of DR, while multi-class classification aims to identify the severity or stage of the disease. The latter approach is more challenging as it requires the model to distinguish between different stages of DR, which can have subtle differences. Several studies have investigated both binary and multi-class classification of DR. In this section, we will review studies that focus on binary classification, studies that focus on multi-class classification, as well as studies that include both binary and multi-class classification of DR.

\subsection{Binary classification of DR}

Machine learning-based models and computer vision methods have been widely utilized in DR identification. As an illustration, Priya et al. \cite{Priya2012} proposed a method based on computer vision for detecting DR in color fundus images. They employed image processing techniques to extract features from the original images, which were then fed into an SVM for DR diagnosis. The method yielded a sensitivity of 98\%, specificity of 96\%, and accuracy of 97.6\% on a test dataset of 250 images. In a study by Asha et al. \cite{Asha2015}, Naïve Bayes (NB), Multi-Layer Perceptron (MLP), and Extreme Learning Machine (ELM) were employed for the identification of DR from a total of 100 fundus images. The results showed that ELM outperformed the other models, achieving an accuracy of 90\%. Another study by Chetoui et al. \cite{Chetoui2018} proposed the utilization of various texture features, such as Local Ternary Pattern (LTP) and Local Energy-based Shape Histogram (LESH), which outperformed Local Binary Pattern (LBP) extracted features. They used SVM to classify the extracted histogram and proposed a histogram binning scheme for feature representation. They validated their approach on the public Messidor \cite{Messidor} database and obtained an accuracy of 90.4\% employing SVM with a Radial Basis Function kernel (SVM-RBF). Ramasamy et al. \cite{Ramasamy2021} utilized textural gray-level features such as co-occurrence and run-length matrix, as well as Ridgelet Transform coefficients to extract features, which were then fused. They employed Sequential Minimal Optimization (SMO) classifier for diagnosing DR. The method achieved 98.87\% sensitivity, 95.24\% specificity, and 97.05\% accuracy on the DIARETDB1 dataset \cite{Kauppi2007}, and 90.9\% sensitivity, 91.0\% specificity, and 91.0\% accuracy on the Kaggle \cite{Dr224} dataset.

Deep convolutional neural networks (CNNs) have demonstrated promising results in the automatic detection of diabetic retinopathy using color fundus images, outperforming most handcrafted feature-based methods in various studies. Xu et al. \cite{Xu2017} achieved an accuracy of 94.5\% using CNN methodology, outperforming some of the classical approaches such as SVM and random forest (RF). They utilized the dataset provided by Kaggle \cite{DrKaggle} and implemented data augmentation techniques such as translation, stretching, rotation, and flipping to increase the size of the dataset and enhance their model’s generalization. Pao et al. \cite{Pao2020} developed a method that utilized entropy calculations of fundus images to improve the detection of lesion edges and identify important regions for diagnosing DR. The proposed bi-channel CNN method attained an accuracy of 87.83\%, a sensitivity of 77.81\%, a specificity of 93.88\%, and an AUC score of 0.93. Das et al. \cite{Das2021} proposed a technique to extract the branching retinal vasculature using maximal principal curvature, adaptive histogram equalization (AHE), and morphological opening. They used a CNN for classification and evaluated their algorithm using the DIARETDB1 dataset, achieving an accuracy of 98.7\% and a precision of 97.2\%. Gargeya et al. \cite{Gargeya2017} introduced a data-driven deep learning algorithm to classify fundus images as healthy or not with high reliability. The model achieved an AUC of 0.97 with a specificity of 98\% and sensitivity of 94\% on 5-fold cross-validation using their local dataset. Furthermore, the model was tested against the independent Messidor-2 and E-Ophtha databases, achieving AUC scores of 0.94 and 0.95, respectively. Alazzam et al. \cite{Alazzam2021} used restricted Boltzmann machines (RBM) and optimal path forest (OPF) models to analyze specialized retinal images and achieved an overall diagnostic accuracy of 89.47\%. The RBM-1000 model had the highest accuracy, and both RBM-1000 and OPF-1000 correctly identified all the instances without DR signs, indicating their high specificity. The RBM-500 model had 100\% sensitivity in detecting DR signals. Zago et al. \cite{Zago2020} suggested a lesion localization model for DR diagnosis, which employs a deep network patch-based approach. The model is trained on the DIARETDB1 dataset and evaluated on several other datasets, including Messidor, achieving an AUC score of 0.912 and a sensitivity of 94\%. The authors used transfer learning with the VGG16 model and initialized it with the pre-trained weights from the ImageNet dataset \cite{Russakovsky2015}, demonstrating the potential of transfer learning in improving performance. 

\subsection{Multi-class classification of DR}

Deep learning (DL) techniques have gained increasing attention in recent years for identifying different stages of DR due to their promising performance compared to traditional ML methods. While ML-based approaches have been used to identify the severity levels of DR, DL techniques have mostly been shown to outperform them in this task.

In 2016, Doshi et al. \cite{Doshi2017} introduced a CNN-based method for grading DR. The proposed model employed a combination of convolutional layers, pooling layers, and fully connected layers to learn patterns and features from retinal images. The method achieved a single model quadratic weighted kappa (QWK) of 0.386, while ensembling three similar models resulted in an improved score of 0.3996. Dondeti et al. \cite{Dondeti2020} used the neural architecture search network (NASNet) to create an efficient feature representation of fundus images. Experimental studies on the benchmark APTOS 2019 dataset demonstrated that deep features obtained from NASNet, when transformed using t-distributed stochastic neighbor embedding (t-SNE), provided a more discriminative representation of fundus images, achieving an accuracy of 77.90\% with $\nu$-SVM classifier. They also achieved a precision of 76\%, recall of 77\%, and F1 score of 75\% for the five-stage grading of DR using their proposed method. They used 80\% of 3662 images for training and 20\% of the data for testing. Kassani et al. \cite{Kassani2019} proposed a modified Xception architecture for the problem. Their method involved deep layer aggregation to combine multilevel features from different convolutional layers, which were then fed into an MLP for training. They also evaluated the performance of their approach using 343 fundus images from the APTOS 2019 dataset and reported a classification accuracy of 83.09\%, sensitivity of 88.24\%, and specificity of 87\%. Gangwar et al. \cite{Gangwar2021} presented a hybrid model for DR stage grading. The approach incorporated transfer learning using pre-trained Inception-ResNet-v2 and extended it with a custom block of CNN layers. The performance of the proposed model was assessed on two datasets, Messidor-1 and APTOS 2019, resulting in test accuracies of 72.33\% and 82.18\%, respectively. Dekhil et al. \cite{Dekhil2019} introduced a CNN architecture composed of a preprocessing stage, followed by five layers of convolution, rectified linear, pooling, and three fully connected layers. Their method attained a testing accuracy of 77\% and a QWK score of 78\% when assessed on 15\% of 3662 images from the APTOS 2019 dataset. Qummar et al. \cite{Qummar2019} employed an ensemble of five CNN models, including ResNet50, Inceptionv3, Xception, DenseNet121, and DenseNet169, and their model achieved an accuracy of 80.8\%, a sensitivity of 51.5\%, and specificity of 86.72\%.

\subsection{Binary and multi-class classification of DR}

Bodapati et al. \cite{Bodapati2020} utilized multiple pre-trained CNN models to extract features from fundus images, which were then blended using a multi-modal fusion module. The proposed approach achieved an accuracy of 96.10\% for DR detection and an accuracy of 80.96\% for DR severity level prediction. They used 80\% of the APTOS 2019 dataset for training and the remaining 20\% for testing their proposed model. Kumar et al. \cite{Kumar2021} proposed a combination of VGG16 and Capsule network models to create a hybrid approach. This resulted in 97.05\% accuracy for identifying DR and 75.50\% accuracy for classifying the five stages, tested on 15\% of the APTOS 2019 dataset. 

More recently, in 2022, Islam et al. \cite{Islam2022} presented a supervised contrastive learning (SCL) method for the detection of DR and its severity stages using the APTOS 2019 dataset. The SCL method is a two-stage training method with a supervised contrastive loss function that overcomes the limitations of traditional deep learning models. The proposed model demonstrated an accuracy of 98.36\% and an AUC of 98.50\% for binary classification of DR. For the five-stage grading task, the model attained an accuracy of 84.36\% and an AUC of 93.81\%. They used 75\% of the APTOS 2019 for training and 15\% for testing purposes.

\section{Materials and methods}

\subsection{Dataset}

Several publicly available datasets are used for the detection and grading of the stages of diabetic retinopathy, such as Messidor, IDRiD, EyePACS, and APTOS 2019. For this research, we selected the APTOS 2019 dataset, which was obtained from the Kaggle competition. This dataset contains fundus images that were provided by the Aravind Eye Hospital in India and have been labeled with one of five categories ranging from 0 to 4, indicating the severity of diabetic retinopathy \cite{Jiaa2015}:
\begin{itemize}
    \item No DR (label 0)
    \item Mild (label 1)
    \item Moderate (label 2)
    \item Severe (label 3)
    \item Proliferative DR (label 4)
\end{itemize}

The dataset comprised 3662 and 1928 training and test samples of varying sizes, respectively. However, the label annotations for the test samples were kept private to evaluate the submissions in the competition. Therefore, the test samples were excluded from this experiment. In this study, we used the publicly available training set of 3662 fundus images with label annotations, which were split into training (70\%), validating (10\%), and testing (20\%) datasets for the purpose of DR diagnosis and stage grading. The distribution of images in the APTOS 2019 dataset for binary and multi-class classification is presented in Table \ref{tbl1}.

\begin{table*}
    \caption{Distribution of images in the APTOS 2019 dataset for binary and multi-class classification.}\label{tbl1}
    \centering
    \begin{tabularx}{\textwidth}{Xcccccc}
        \toprule
        \multirow{2}{*}{Classification} & \multirow{2}{*}{DR Stage} & \multirow{2}{*}{Number of images} & \multicolumn{3}{c}{Proportion of images} & \\
        \cmidrule{4-6}
        & & & Train (70\%) & Validation (10\%) & Test (20\%) & \\
        \midrule
        \textbf{Binary} & No DR (0) & 1805 & 1264 & 180 & 361 & \\
        & DR (1) & 1857 & 1303 & 184 & 370 & \\
        \textbf{Multi-class} & No DR (0) & 1805 & 1264 & 180 & 361 & \\
        & Mild (1) & 370 & 259 & 37 & 74 & \\
        & Moderate (2) & 999 & 701 & 99 & 199 & \\
        & Severe (3) & 193 & 136 & 19 & 38 & \\
        & PDR (4) & 295 & 207 & 29 & 59 & \\
        \bottomrule
    \end{tabularx}
\end{table*}

\begin{figure}
    \centering
    \includegraphics[width=\linewidth]{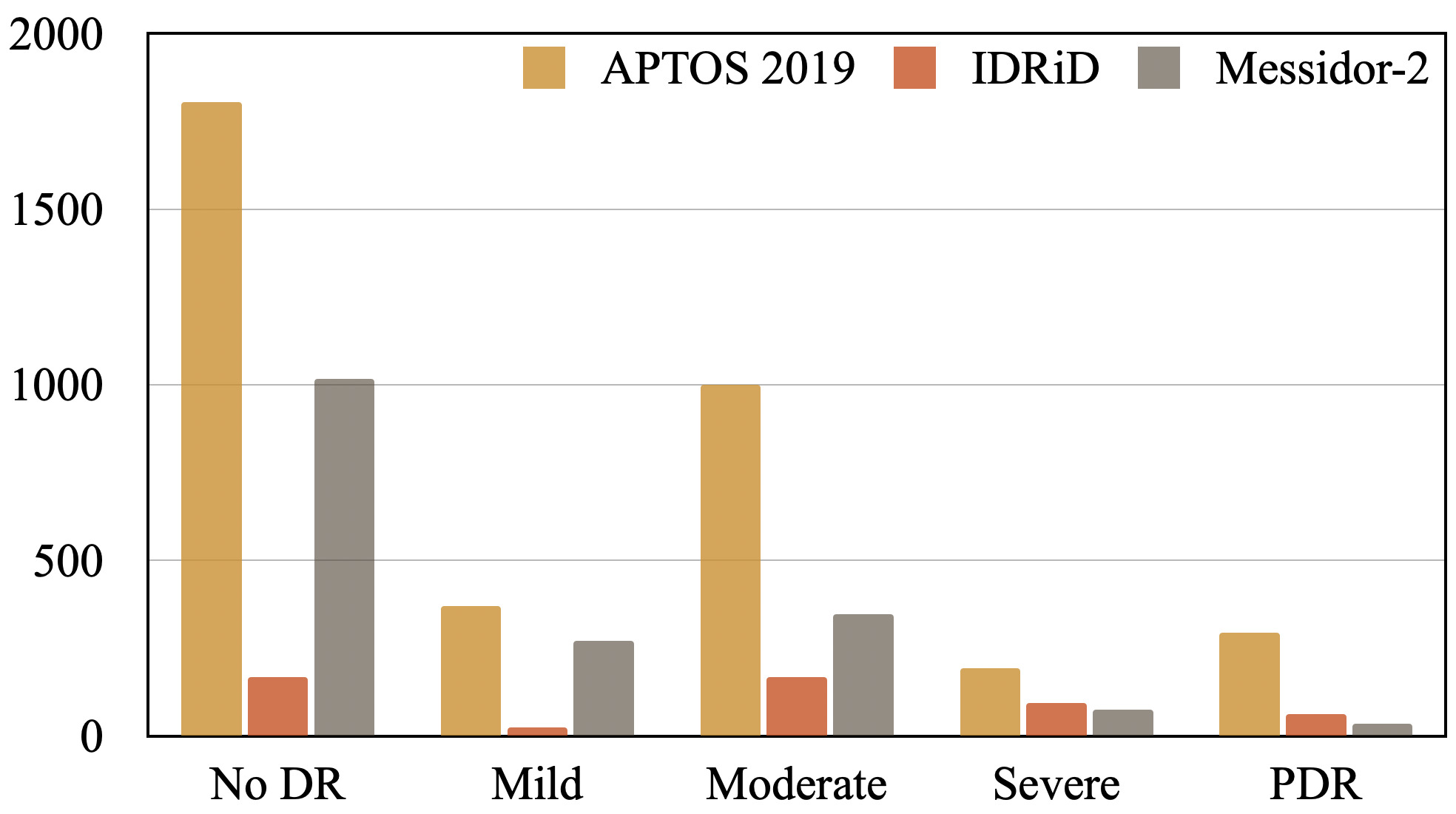}
    \caption{The distribution of classes in APTOS 2019, IDRiD, and Messidor-2 datasets.}
    \label{fig:fig3}
\end{figure}

To enhance the performance of our model during the training phase, we merged some categories of Messidor-2 \cite{Krause2018} and IDRiD \cite{Porwal2018} datasets with the APTOS 2019 dataset for training our model, as described in detail in Subsection \ref{sec:merged}. The Messidor-2 dataset contains 1744 fundus images, including 727 with diabetic retinopathy and 1017 without the disease, along with grading annotations ranging from 0 to 4. We used the relabeled version of Messidor-2, in which the grades were adjudicated by a panel of three retina specialists \cite{Krause2018}. The IDRiD dataset, on the other hand, consists of 516 fundus images, along with corresponding annotations. By utilizing some categories from these datasets, our model was able to learn a more diverse range of features and generalize better to unseen data. The distribution of classes in APTOS 2019, IDRiD, and Messidor-2 datasets are compared in Fig.~\ref{fig:fig3}. Notably, all these three publicly available resources display a similar pattern of imbalanced class distribution.

\subsection{Proposed approach}

Our proposed approach utilizes transfer learning, building on the success of previous studies in this field. To take advantage of this approach, we fine-tuned two popular pre-trained models, ResNet50 and EfficientNetB0, that have shown remarkable results in various image recognition tasks. These models were originally trained on the ImageNet dataset \cite{Russakovsky2015}, which consists of millions of labeled images, enabling them to learn highly informative and discriminative features from images. We selected these models based on their ability to handle complex and diverse image data, which is critical in the case of DR classification. As DR can manifest in different forms, such as microaneurysms, hemorrhages, and exudates, it is crucial to have models that can detect subtle differences in these features. By leveraging these pre-trained models, we aimed to improve the performance of our model in detecting diabetic retinopathy stages.

\begin{figure*}
    \centering
    \includegraphics[width=\linewidth]{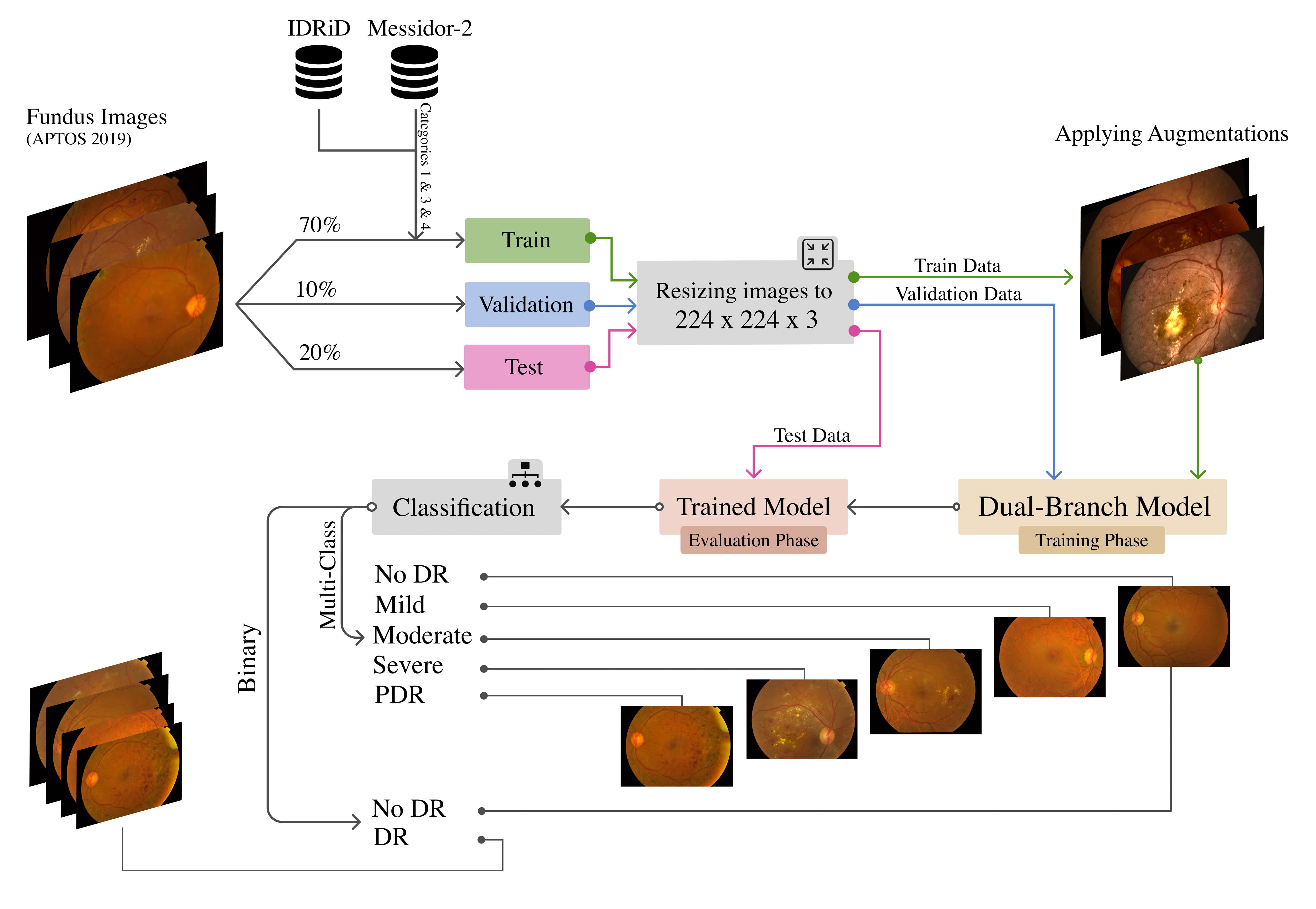}
    \caption{Proposed framework for detection and stage grading of diabetic retinopathy.}
    \label{fig:fig4}
\end{figure*}

To mitigate the issue of imbalanced distribution in available datasets for DR detection and grading, we explored various approaches to modify the training dataset. Our findings indicated that merging certain categories of three publicly available datasets during training yielded the best performance. Additionally, to mitigate the effects of imbalanced data further, we employed a specialized loss function during the training process. These efforts enabled our model to achieve superior performance across all DR classes, even in the presence of class imbalance. The overall proposed framework is depicted in Fig.~\ref{fig:fig4}.

\subsection{Pre-trained models}

\begin{figure}
    \centering
    \includegraphics[width=\linewidth]{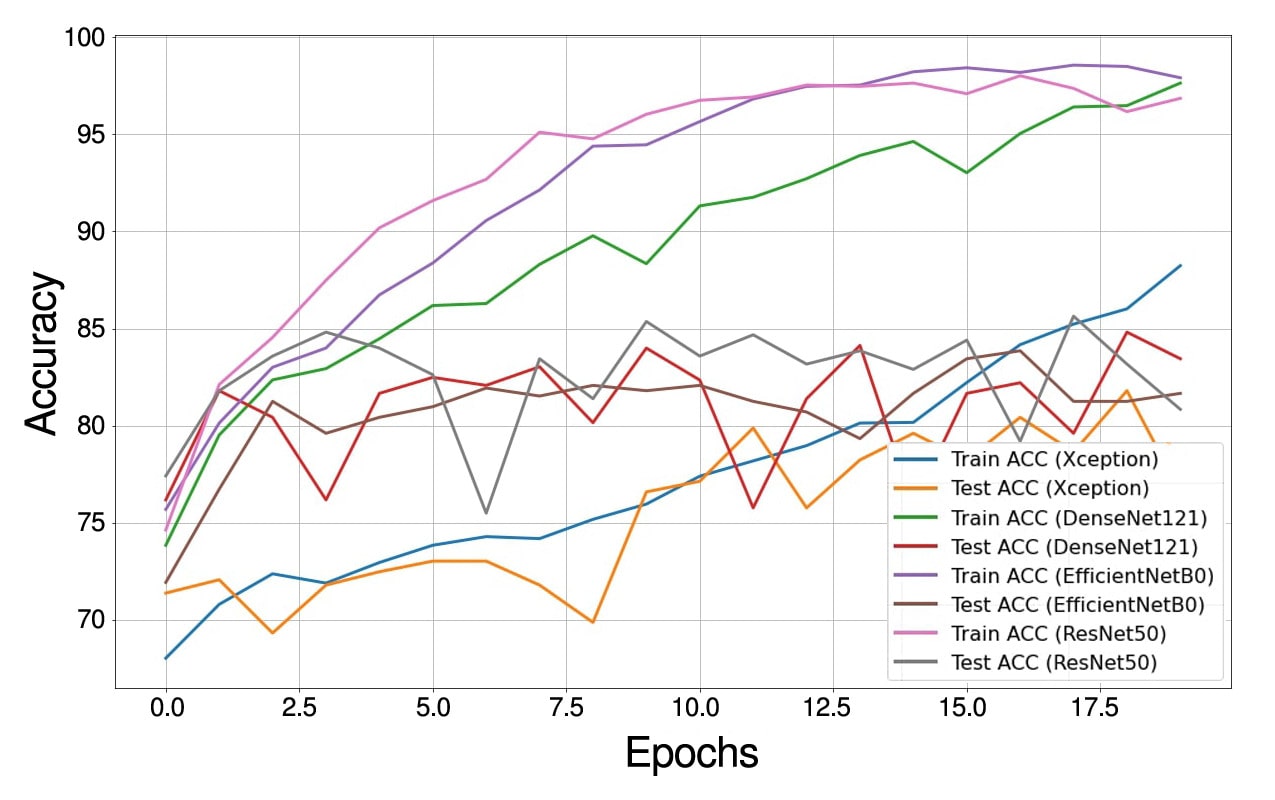}
    \caption{Training and test accuracy of pre-trained models with APTOS 2019 dataset for DR stage grading.}
    \label{fig:fig5}
\end{figure}

Choosing an appropriate pre-trained model as the feature extractor is a crucial step in transfer learning-based approaches for image classification tasks. We experimented with four popular pre-trained models, ResNet50, Xception, DenseNet121, and EfficientNetB0, which have shown remarkable success in various image recognition tasks. To assess the performance of these models, we trained and evaluated them on 80\% and 20\% of the APTOS 2019 dataset, respectively. The results of these experiments, including the training and test accuracies obtained with each model, are presented in Fig.~\ref{fig:fig5}. The accuracy results depicted in the figure indicate that ResNet50 and EfficientNetB0 outperformed the other models, providing sufficient justification for their selection as our feature extractors.

\subsubsection{ResNet50}

ResNet50 \cite{He2016}, introduced in 2015, is a deep CNN that employs residual learning and consists of 50 layers. Residual learning involves adding shortcut connections to the network, allowing information to be passed through the network without being affected by the layers in between. This allows for easier training of very deep networks, as it reduces the vanishing gradient problem. The architecture has shown remarkable performance in various computer vision tasks.

ResNet50 was selected as one of our candidate feature extractors because of its high accuracy and ability to learn highly informative features from images.

\subsubsection{Xception}

Xception \cite{Chollet2017} is a deep CNN that was introduced in 2017. The network is based on the Inception \cite{Szegedy2015} architecture, but instead of using traditional convolutional layers, it uses depthwise separable convolutional layers. Depthwise separable convolutional layers consist of a depthwise convolution layer followed by a pointwise convolution layer. This allows the network to learn more diverse and complex features with fewer parameters than traditional CNNs.

We chose Xception as one of our candidate feature extractors due to its high performance on various image recognition tasks and its proven success in handling complex image data. The use of depthwise separable convolutional layers makes Xception an efficient choice for image classification, as it can learn more complex features while using fewer parameters.

\subsubsection{DenseNet121}

DenseNet121 \cite{Huang2017} is a deep CNN that was introduced in 2017. It is based on the idea of dense connectivity, where each layer is connected to every other layer in a feed-forward fashion. This connectivity pattern allows DenseNet121 to learn highly discriminative features from images, as each layer has access to the feature maps of all preceding layers.

We chose DenseNet121 as one of our candidate feature extractors due to its high accuracy and compact architecture, which enables it to learn complex features with fewer parameters than most other architectures.

\subsubsection{EfficientNetB0}

EfficientNetB0 \cite{Tan2019} is a deep CNN that was introduced in 2019. It was designed to have a higher capacity for learning complex features from images while using fewer parameters than other architectures. EfficientNetB0 achieves this by using a compound scaling method that optimizes the scaling of depth, width, and resolution of the network. The EfficientNet family includes several variations, with EfficientNetB0 being the smallest and most computationally efficient.

EfficientNetB0 was selected as one of our candidate feature extractors due to its superior performance in various image recognition tasks and its efficiency in terms of model size and computational resources.

\subsection{Preprocessing}

In the preprocessing stage, we applied various image augmentations to improve the quality of fundus images and increase the size and diversity of our dataset. The following section will provide more details on the specific augmentations we utilized. We also resized all the images to a uniform size of 224 x 224 x 3 to ensure compatibility with the input size requirement of the selected deep learning models, as well as to address the issue of varying image sizes in the original dataset.

\subsection{Data augmentation} \label{sec:augmentation}

\begin{figure}
    \centering
    \includegraphics[width=\linewidth]{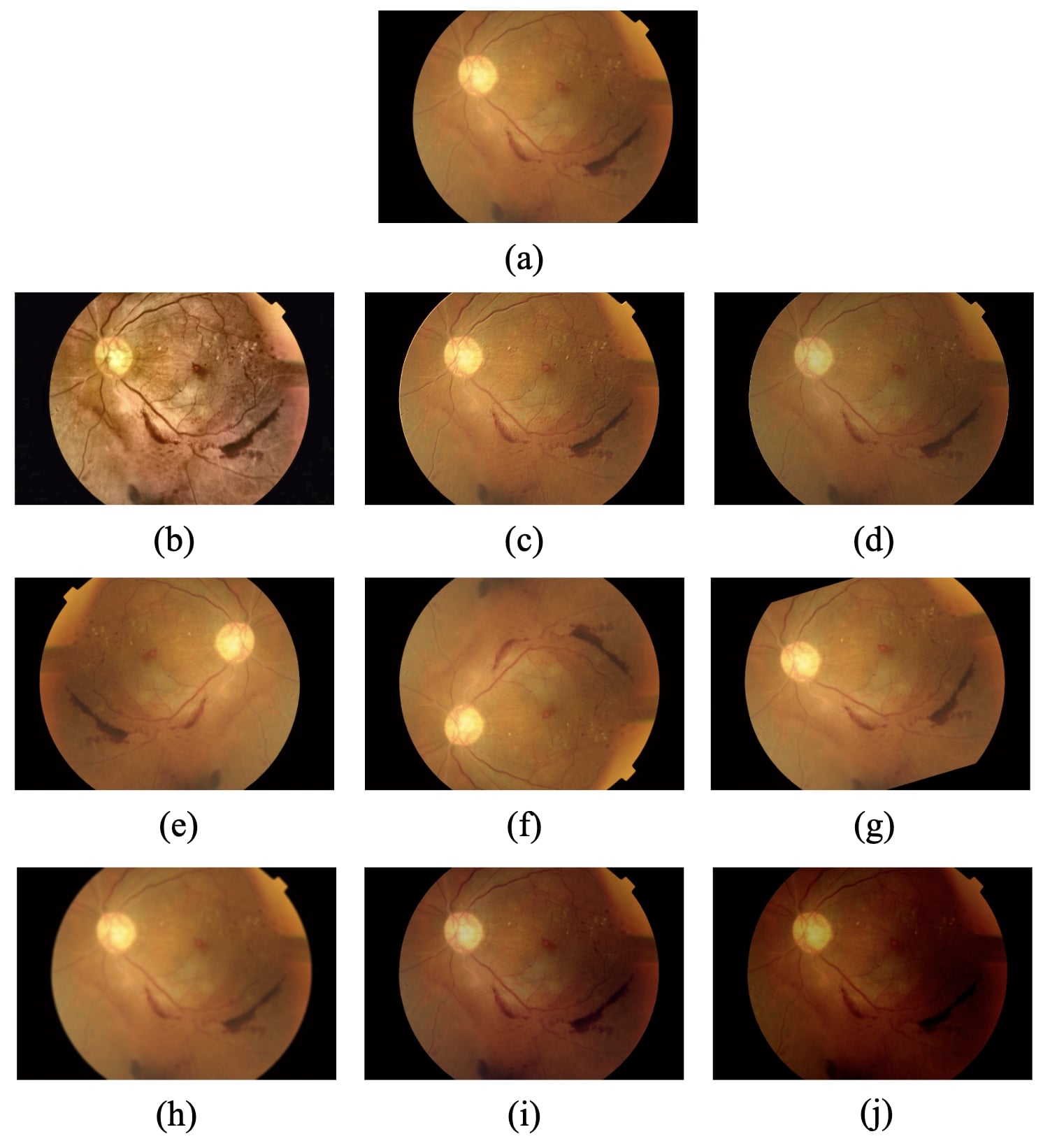}
    \caption{Applied augmentations for a (a) fundus image: (b) CLAHE, (c) emboss, (d) sharpen, (e) horizontal flip, (f) vertical flip, (g) random rotation, (h) blur, (i) fancy PCA, (j) random brightness and contrast.}
    \label{fig:fig6}
\end{figure}

Various image augmentation techniques were employed to enhance the quality of the images and increase the size and diversity of the dataset. Specifically, the Albumentations library \cite{Buslaev2020} was utilized to apply blur, vertical flip, horizontal flip, random rotation, sharpening, CLAHE (Contrast Limited Adaptive Histogram Equalization), embossing, fancy PCA, and random brightness and contrast to the fundus images. All of the mentioned augmentations are shown for a fundus image in Fig.~\ref{fig:fig6}.

\subsection{Network architecture}

The proposed model for detecting and grading different stages of diabetic retinopathy is a deep dual-branch model. Using multiple branches is a common technique in deep learning to combine the strengths of multiple models or feature extractors to achieve improved performance.

Our suggested network consists of two high-performing ImageNet-pre-trained models, ResNet50 \cite{He2016} and EfficientNetB0 \cite{Tan2019}, and several layers, including max pooling, batch normalization, and fully connected layers.

The input to the model is a fundus retinal image passed through the ResNet50 model to extract features. The output from ResNet50 is then passed through a max pooling layer to reduce the spatial dimensions of the output. The resulting output is then passed through a rectified linear unit (ReLU) activation function to introduce non-linearity.

The exact process is repeated for the EfficientNetB0 model, and the resulting output is passed through the same max pooling, ReLU, and batch normalization layers as the output from ResNet50. The outputs from both models are then concatenated. 

\begin{figure*}
    \centering
    \includegraphics[width=\linewidth]{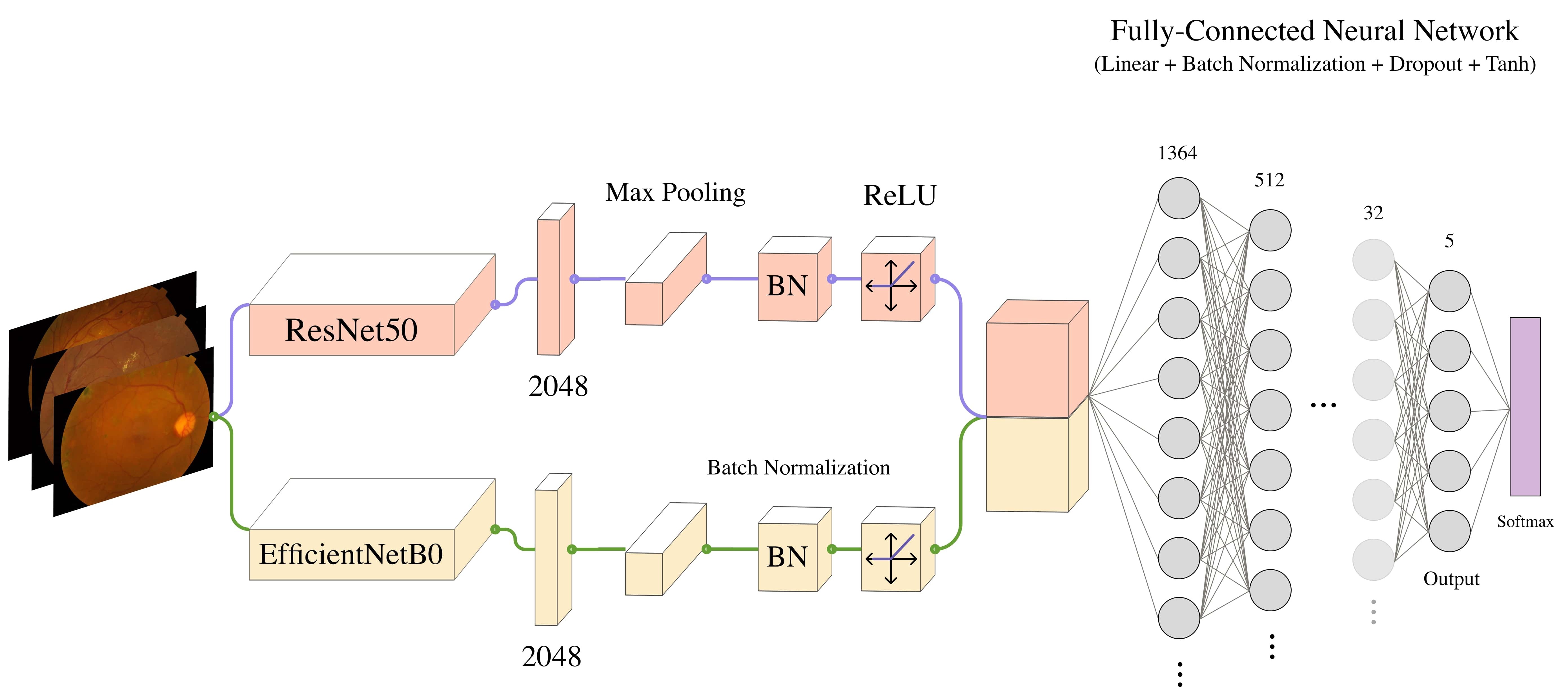}
    \caption{Architecture of the dual-branch proposed model for diabetic retinopathy detection and severity grading.}
    \label{fig:fig7}
\end{figure*}

The concatenated output is then passed through a fully connected neural network consisting of four layers of linear transformations, batch normalization, activation functions, and dropout regularization to prevent overfitting. The output of the final linear layer is the predicted class probabilities for each of the diabetic retinopathy stages. The proposed network structure is shown in Fig.~\ref{fig:fig7}.

\subsection{Loss function}

To address the issue of imbalanced data, we chose to use the complement cross entropy (CCE) loss function, specifically designed to give more importance to the minority class samples. CCE neutralizes the probabilities of the incorrect classes and ensures that the ground truth class has a higher softmax probability than the incorrect classes \cite{Kim2021}. As a result, it is expected to deliver classification results that are more accurate and robust for imbalanced distributions, making it a suitable choice for our DR classification problem.

\section{Experimental results and analysis}
In this study, we conducted experiments on the APTOS 2019 dataset and a merged dataset consisting of APTOS 2019, Messidor-2, and IDRiD datasets. Our models were trained using the PyTorch framework on an NVIDIA GeForce RTX 3070i GPU, employing the stochastic gradient descent (SGD) optimizer. Throughout the training process, we saved the weights corresponding to the best performance based on the quadratic weighted kappa (QWK) score on the validation set. These saved weights were then utilized to initialize the models for making predictions on new images.

To ensure consistency with established literature, we designed the test set to encompass 20\% of each category within the original APTOS 2019 dataset. This allowed us to compare our results with previous studies that employed a similar fraction of the APTOS 2019 for evaluation purposes. Notably, we used the same test data in all of our experiments, ensuring consistent evaluations.

\subsection{Evaluation metrics}

The imbalanced dataset poses a challenge when assessing performance using accuracy. This is because accuracy tends to prioritize the majority class, assigning more weight to its correct predictions while potentially overlooking the minority class. \cite{Akosa2017}. In this situation, it is recommended to use the quadratic weighted kappa (QWK) score as the primary evaluation metric. QWK measures inter-rater agreement in multi-class classification problems by comparing the expected and predicted scores, with a score range of -1 (complete disagreement) to 1 (complete agreement). Higher scores indicate better agreement between the raters or better model performance \cite{Cohen1968}. $\kappa$ is defined as:

\begin{equation}
\label{eq:qwk}
  \kappa = 1- \frac{\sum_{i = 1}^k\sum_{j = 1}^k W_{ij}O_{ij}}{\sum_{i = 1}^k\sum_{j = 1}^k W_{ij}E_{ij}}
\end{equation}

$k$ represents the number of output classes, $O_{ij}$ and $E_{ij}$ correspond to elements in the observed and expected matrices, respectively. The calculation of $W_{ij}$ is defined as follows:
\begin{equation}
\label{eq:w}
  W_{ij} = \frac{(i - j)^2}{(k - 1)^2}
\end{equation}

Accuracy is a metric that quantifies the ratio of correct predictions made by the model to the total number of predictions (Equation \ref{eq:acc}). A higher accuracy indicates that more predictions were correct.
\begin{equation}
\label{eq:acc}
  Accuracy = \frac{TP+TN}{TP+FP+TN+FN}
\end{equation}

The variables TP, TN, FP, and FN are defined as True Positive, True Negative, False Positive, and False Negative, respectively. Precision, as indicated in Equation (\ref{eq:pr}), measures the ratio of TPs to the total number of positive predictions.
\begin{equation}
\label{eq:pr}
  Precision = \frac{TP}{TP+FP}
\end{equation}

Sensitivity, also known as recall, measures the proportion of actual positive instances that are correctly identified by the model (Equation \ref{eq:re}).
\begin{equation}
\label{eq:re}
  Sensitivity = \frac{TP}{TP+FN}
\end{equation}

Specificity measures the proportion of actual negative instances that are correctly identified by the model (Equation \ref{eq:sp}).
\begin{equation}
\label{eq:sp}
  Specificity = \frac{TN}{TN+FN}
\end{equation}

F1-score, as defined by Equation \ref{eq:f1}, is calculated as the harmonic mean of precision and sensitivity, and ranges from 0 to 1, where higher scores indicate superior performance.
\begin{equation}
\label{eq:f1}
  F1-Score = 2\times\frac{Precision \times Sensitivity}{Precision + Sensitivity}
\end{equation}

We used weighted averaging to evaluate metrics that require averaging in multi-class classification, such as precision, sensitivity, specificity, and F1-score. This approach accounts for the contribution of each class to the overall model performance, avoiding bias towards larger or smaller classes. It provides a more accurate assessment of the model's effectiveness in detecting and grading DR.

\subsection{Experiments on APTOS 2019}\label{sec:aptos-only}
We employed a range of augmentation techniques, outlined in Subsection \ref{sec:augmentation}, to augment the training images across all categories in the APTOS 2019 dataset. This resulted in a tenfold increase in our final training dataset, growing from 2567 to 25670 instances. Our model was trained for 30 epochs, with a training duration of 2 hours and 32 minutes, and delivered state-of-the-art performance, surpassing many existing works in the literature. 

\begin{figure*}
    \centering
    \includegraphics[width=\linewidth]{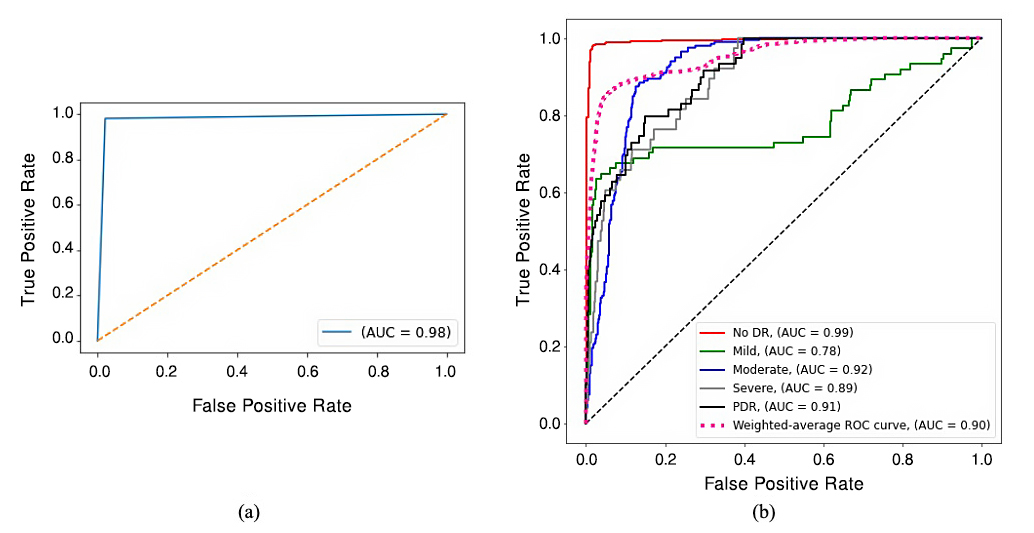}
    \caption{ROC curve for (a) binary and (b) multi-class classification with APTOS 2019 (trained on augmented APTOS 2019)}
    \label{fig:fig8}
\end{figure*}

The effectiveness of our model was further demonstrated through the receiver operating characteristics (ROC) curve analysis. The ROC curve summarizes the classifier's performance in distinguishing between classes, and the area under the curve (AUC) score serves as a measure of this capability. Fig.~\ref{fig:fig8} displays the ROC curves for both binary and multi-class classification of our model, which was trained using the APTOS 2019 dataset with augmented images. Fig.~\ref{fig:fig8}(a) illustrates an AUC score of 98\% for binary classification, and Fig.~\ref{fig:fig8}(b) demonstrates an AUC score of 90\% for multiclass classification. Additionally, the AUC scores for the individual stages, namely normal, mild, moderate, severe, and PDR, are 99\%, 78\%, 92\%, 89\%, and 91\%, respectively.

\begin{figure*}
    \centering
    \includegraphics[width=\linewidth]{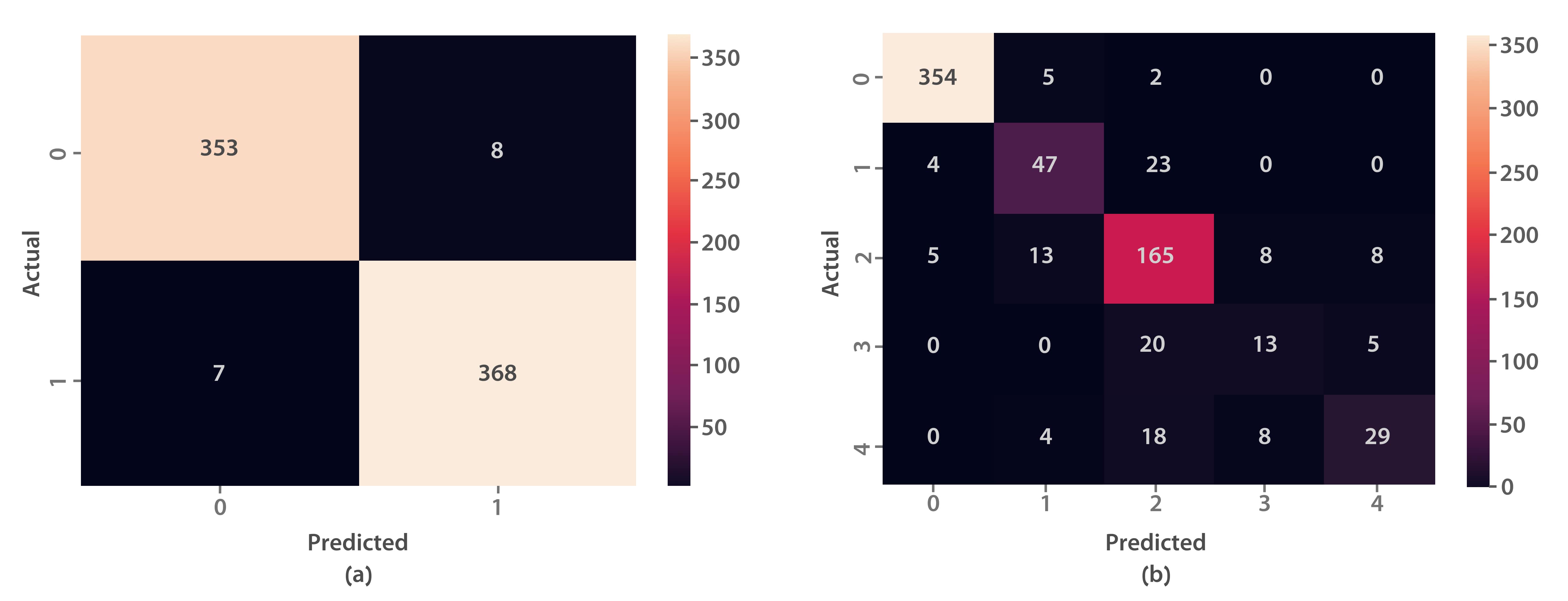}
    \caption{Confusion matrix for (a) binary and (b) multi-class classification with APTOS 2019 (trained on augmented APTOS 2019)}
    \label{fig:fig9}
\end{figure*}

We report the overall performance of our model for binary and multi-class classification in Table \ref{tbl2}, which compares the results of training our model using the original APTOS 2019 and the APTOS 2019 with the application of augmentations. The table shows that applying augmentations led to better performance in all evaluation metrics, demonstrating the effectiveness of the augmentation techniques. The confusion matrices for binary and multi-class classification of the model are shown in Fig.~\ref{fig:fig9}.

\subsection{Experiments on merged dataset} \label{sec:merged}
In this section, we describe our methodology for merging datasets and selecting categories to create a new dataset for training our model. We also present our findings from experiments conducted on this new multi-center dataset.

For our experiments, we used the APTOS 2019 dataset as our main source of data. However, we also wanted to incorporate additional data to improve our model's performance. Initially, we attempted to merge the APTOS 2019 dataset with two other datasets, Messidor-2 and IDRiD. After merging the APTOS 2019 dataset with Messidor-2 and IDRiD, we applied image augmentations to all the categories. This step was taken to enhance the performance of our model further. This approach resulted in remarkable achievements, with our model achieving an overall accuracy of 88.65\% and a QWK score of 90\% for multi-class classification. These impressive results highlight the significant impact of integrating diverse datasets and employing appropriate data augmentation techniques in enhancing the accuracy and robustness of our model.

\begin{figure}
    \centering
    \includegraphics[width=\linewidth]{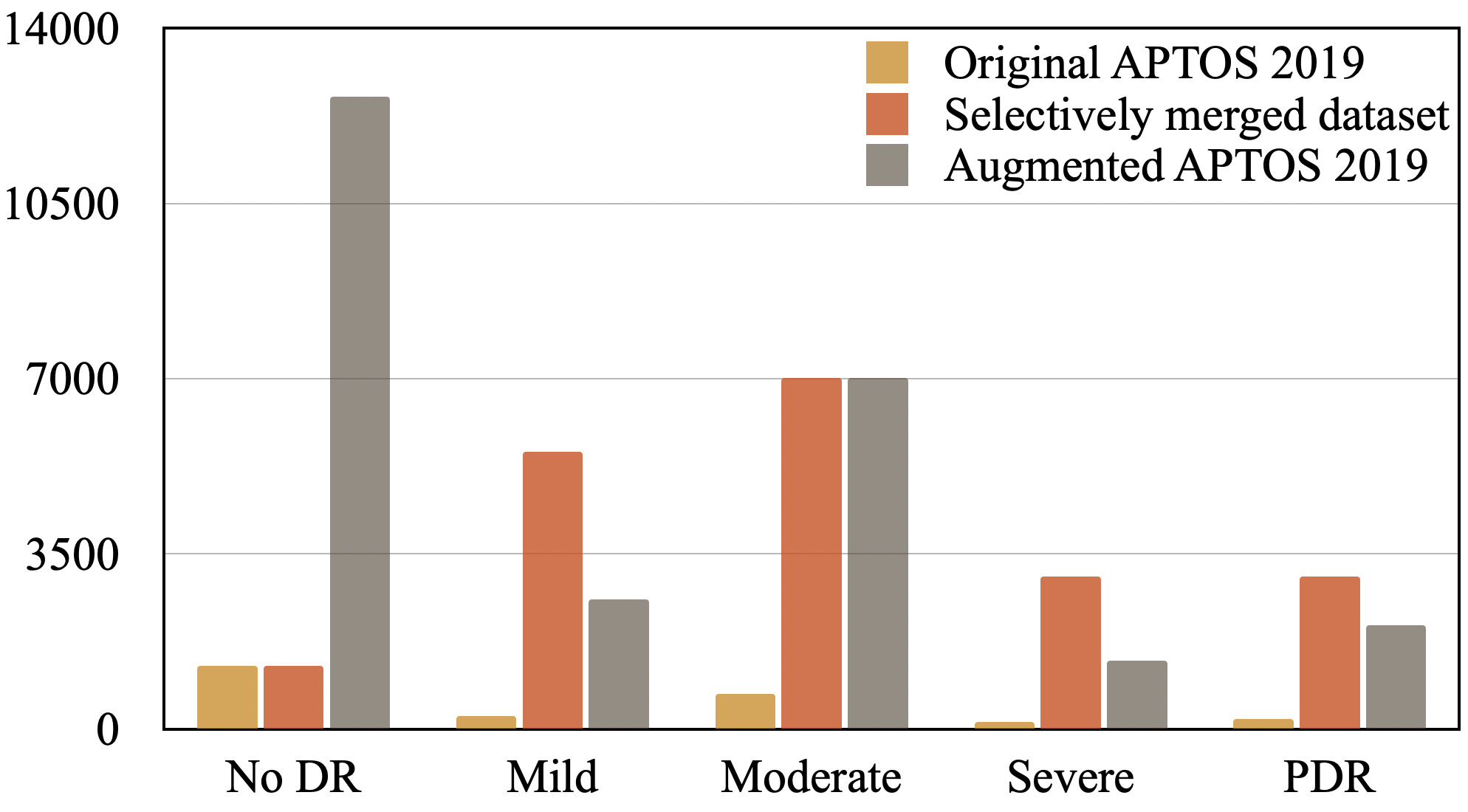}
    \caption{Class distribution comparison between the original APTOS 2019, APTOS 2019 with augmentations, and the selectively merged datasets.}
    \label{fig:fig10}
\end{figure}

With the goal of optimizing efficiency, we made a deliberate decision to merge only categories 1 (mild), 3 (severe), and 4 (PDR) of Messidor-2 and IDRiD with the APTOS 2019 dataset, which had a relatively smaller number of images compared to other categories. Additionally, we strategically chose to apply augmentations only to categories 1 to 4, as category 0 had a sufficient amount of data and was significantly larger than other categories. Furthermore, the evaluation results indicated that the models did not encounter difficulties in identifying this category. Fig.~\ref{fig:fig10} illustrates the distribution of classes in the original APTOS 2019 dataset, APTOS 2019 dataset with augmentations across all classes, and the selectively merged dataset.

This efficient approach not only reduced the data and computation cost but also yielded exceptional performance. The selectively merged dataset, along with the targeted augmentations, outperformed the previous method in terms of all evaluation metrics for multi-class classification. The comparison of models trained with the merged dataset, including all categories and the selectively merged dataset with categories 1, 3, and 4 is presented in Table \ref{tbl3} for multi-class classification. 

\begin{figure*}
    \centering
    \includegraphics[width=\linewidth]{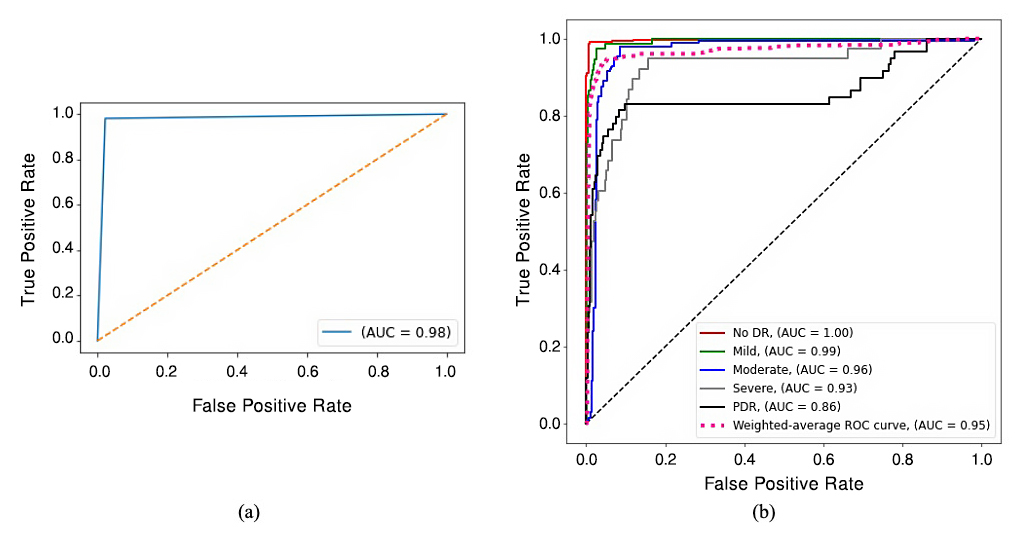}
    \caption{ROC curve for (a) binary and (b) multi-class classification with APTOS 2019 (trained on selectively merged dataset)}
    \label{fig:fig11}
\end{figure*}

\begin{figure*}
    \centering
    \includegraphics[width=\linewidth]{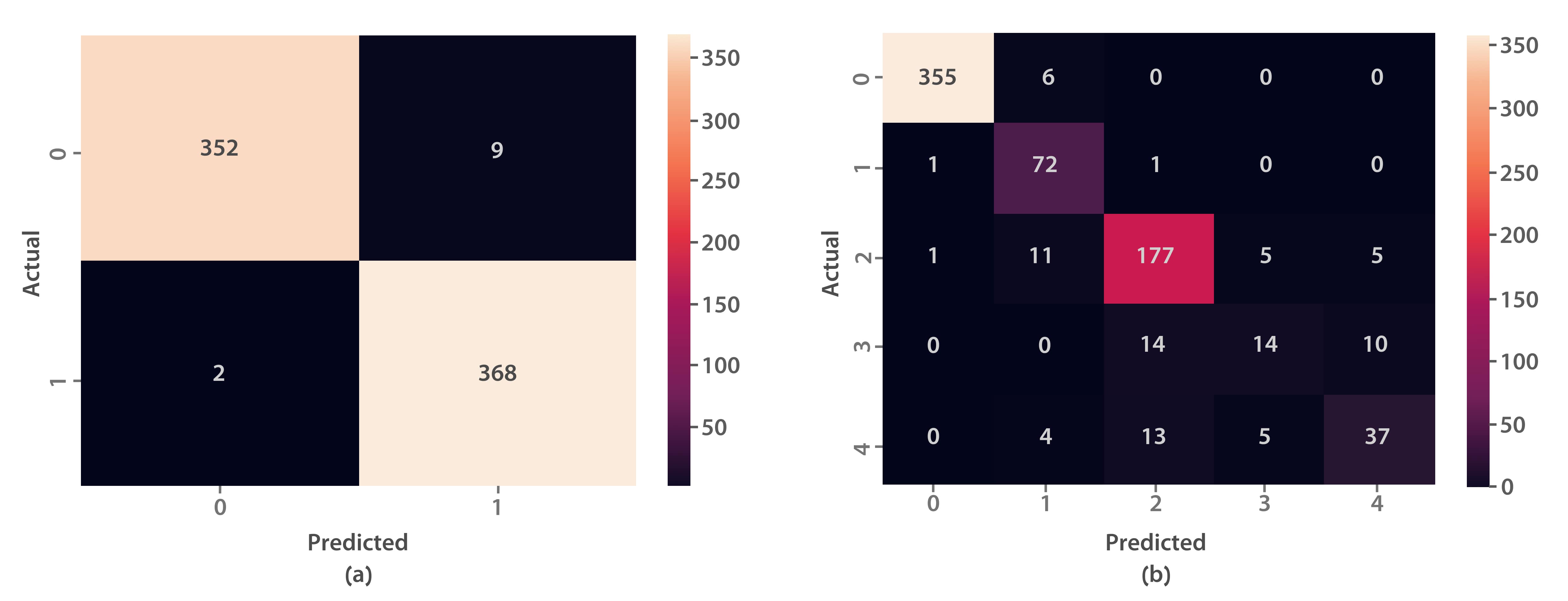}
    \caption{Confusion matrix for (a) binary and (b) multi-class classification with APTOS 2019 (trained on selectively merged dataset)}
    \label{fig:fig12}
\end{figure*}

The resulting model achieved remarkable performance in multi-class classification, with an accuracy of 89.60\%, a QWK score of 93.00\%, an F1-score of 89.15\%, a sensitivity of 89.60\%, and a specificity of 97.72\%. These outstanding results signify the effectiveness of our approach, surpassing the performance of existing methods reported in the literature. Fig.~\ref{fig:fig11} illustrates the impressive AUC scores obtained by the proposed model, with 98\% for binary classification and 95\% for multi-class classification. The corresponding confusion matrices for binary and multi-class classification of the model can be found in Fig.~\ref{fig:fig12}. Additionally, Table \ref{tbl4} presents the class-wise evaluation of the proposed model for both binary and multi-class classification.
\begin{table}
    \caption{Performance comparison of binary and multi-class classification using the APTOS 2019 dataset with and without augmentations.}\label{tbl2}
    \centering
    \begin{tabularx}{\columnwidth}{Xcccc}
        \toprule
        \multirow{2}{*}{Metrics (\%)} & \multicolumn{2}{c}{w/o augmentations} & \multicolumn{2}{c}{w/ augmentations}\\
        \cmidrule{2-3} \cmidrule{4-5}
        & Binary & Multi-class & Binary & Multi-class\\
        \midrule
        QWK & 92.89 & 87.63 & \textbf{95.90} & \textbf{89.03} \\
        Accuracy & 96.44 & 80.44 & \textbf{97.95} & \textbf{83.17} \\
        Precision & 97.78 & 79.50 & \textbf{97.84} & \textbf{82.66} \\
        Sensitivity & 95.14 & 80.44 & \textbf{98.11} & \textbf{83.17} \\
        Specificity & 97.78 & 93.91 & \textbf{97.78} & \textbf{94.96} \\
        F-1 Score & 96.44 & 79.12 & \textbf{97.98} & \textbf{82.64} \\
        AUC & 96.46 & 85.28 & \textbf{97.94} & \textbf{89.84} \\
        \bottomrule
    \end{tabularx}
\end{table}

\begin{table}
    \caption{Performance comparison of models trained on the merged dataset including all categories and the selectively merged dataset for multi-class classification.}\label{tbl3}
    \centering
    \begin{tabularx}{\columnwidth}{Xcc}
        \toprule
        \multirow{2}{*}{Metrics (\%)} & \multicolumn{2}{c}{Merged dataset} \\
        \cmidrule{2-3}
        & All categories & Selectively merged \\
        \midrule
        QWK & 91.67 &\textbf{93.00} \\
        Accuracy & 88.65 & \textbf{89.60} \\
        Precision & 88.71 & \textbf{89.23} \\
        Sensitivity & 88.65 & \textbf{89.60} \\
        Specificity & 96.41 & \textbf{97.72} \\
        F-1 Score & 87.76 & \textbf{89.15} \\
        AUC & 92.46 & \textbf{94.81} \\
        \bottomrule
    \end{tabularx}
\end{table}

\begin{table*}[b]
 \caption{Class-wise performance evaluation of the proposed model trained on the selectively merged dataset with APTOS 2019.}\label{tbl4}
 \begin{tabularx}{\textwidth}{Xcccccc}
 \toprule
 Classification & QWK (\%) & Accuracy (\%) & Precision (\%) & Sensitivity (\%) & Specificity (\%) & F1-score (\%)\\
 \midrule
    \textbf{Binary classification} &  &  &  &  &  & \\
    No DR & - & - & 99.44 & 97.50 & 99.46 & 98.46\\
    DR & - & - & 97.61  & 99.46 & 97.50 & 98.53\\
    Overall & 96.99 & 98.50 & 97.61 & 99.46 & 97.51 & 98.53\\
    \textbf{Multi-class classification} &  &  &  &  &  & \\
    No DR & - & - & 99.44  & 98.24 & 99.34 & 98.84\\
    Mild & - & - & 77.42  & 97.30 & 96.52 & 86.23\\
    Moderate & - & - & 86.34 & 88.95 & 94.47 & 87.63\\
    Severe & - & - & 58.33 & 36.84 & 98.37 & 45.16\\
    PDR & - & - & 71.15 & 62.71 & 97.63 & 66.66\\
    Overall & 93.00 & 89.60 & 89.23 & 89.60 & 97.72 & 89.15\\
\bottomrule
\end{tabularx}
\end{table*}

\subsection{Hyperparameter tuning}
Bayesian optimization with the Ray Tune library, utilizing a Gaussian Process, was employed to optimize the performance of our proposed model for DR detection and stage grading. The process involved tuning the learning rate and momentum hyperparameters. Bayesian optimization intelligently explored the hyperparameter space, prioritizing regions with promising results from previous iterations. The optimal learning rate and momentum values were determined to be 0.06 and 0.66, respectively. Leveraging these tuned hyperparameters, the proposed model demonstrated enhanced performance in detecting and grading DR, contributing to early diagnosis and treatment of the disease.

\subsection{Regularization}
To prevent overfitting during training, we incorporated dropout layers with a rate of 0.25 after each linear layer in the fully connected neural network of our proposed model. Dropout randomly sets a fraction of the input units to zero during training, which can prevent the network from relying too heavily on specific features and help prevent overfitting \cite{Srivastava2014}. Additionally, we used data augmentation techniques to further increase our model's generalization ability.  Specifically, we applied random rotations, horizontal and vertical flips, brightness adjustments, etc., to the training images to generate new variations of the same image. These techniques help the model learn to recognize the relevant features despite changes in the image's appearance due to factors such as camera angle or brightness. These regularization approaches allowed us to perform well on the test set while avoiding overfitting the training data.

\section{Discussion}\label{sec:discussion}
Detecting diabetic retinopathy in its early stages is of utmost importance as it enables timely interventions and can prevent irreversible vision loss in patients. Our proposed model demonstrated exceptional performance in detecting mild and moderate cases of DR. By accurately identifying these cases, our model can aid in early intervention and treatment, potentially preventing the progression of the disease and reducing the risk of blindness in patients.

\begin{figure*}
    \centering
    \includegraphics[width=\linewidth]{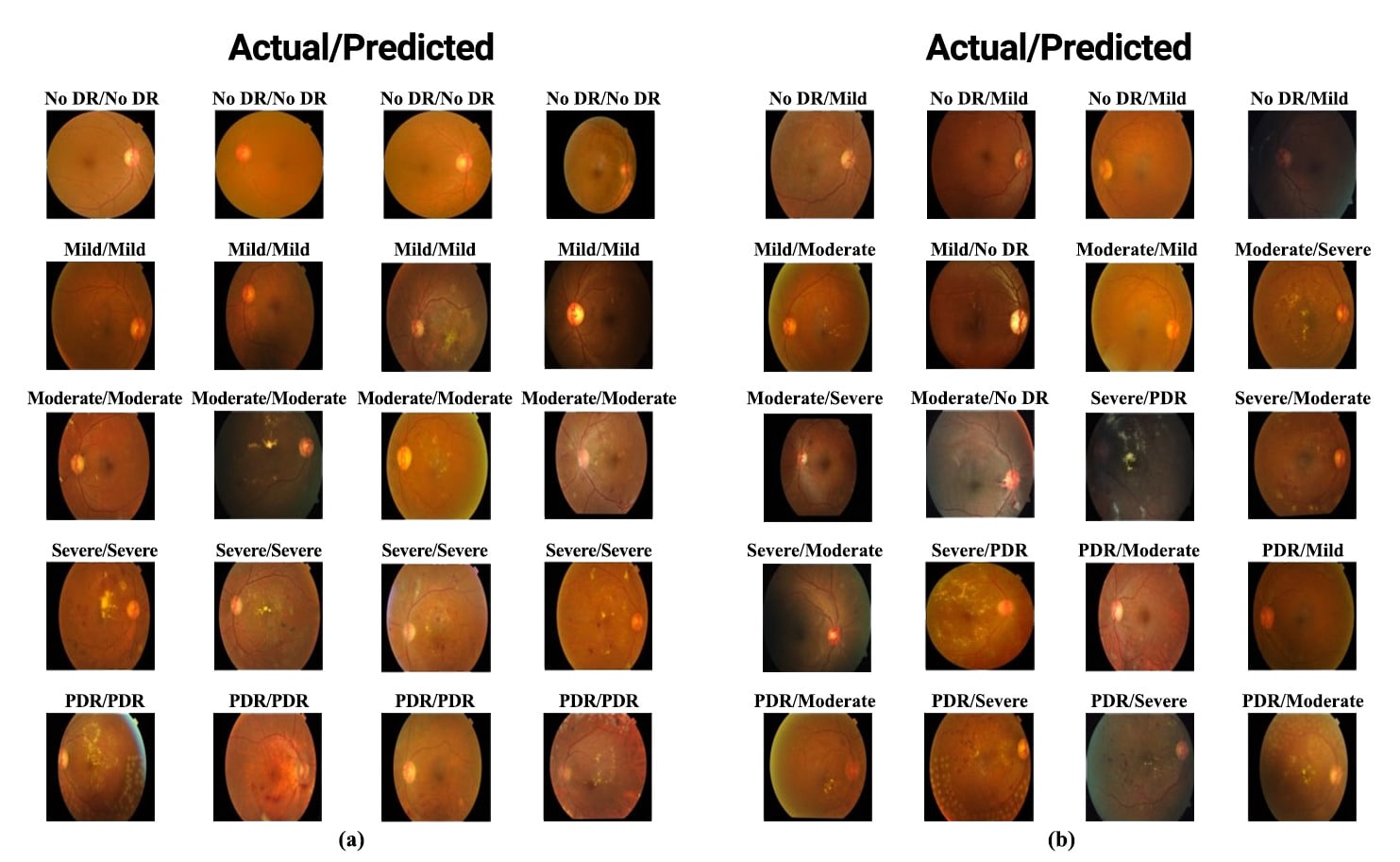}
    \caption{Examples of model predictions on the test dataset, showing both (a) correct classifications and (b) misclassifications.}
    \label{fig:fig13}
\end{figure*}

While our proposed model shows promise in detecting and grading DR, we must acknowledge its limitations to provide a holistic view of its performance. The model displayed limitations distinguishing between severe and proliferative diabetic retinopathy (PDR) cases due to their visually similar characteristics, which even experienced human graders may struggle to differentiate. Furthermore, the severity of DR can vary within each category, adding complexity to accurate classification. Notably, a considerable number of misclassifications in the test dataset were identified as moderate cases, where the presence of exudates in fundus images further complicated the model's performance. Exudates, which are yellowish or white lesions in the retina, were observed in these three cases (moderate, severe, and PDR), potentially contributing to misclassifications and posing challenges for the model in accurately classifying DR stages, particularly in severe cases where its performance was comparatively poorer. Also, our approach exhibits a data-hungry characteristic, relying on a substantial number of images for effective training. While this contributes to the model's performance, it poses challenges in terms of dataset curation and collection. The need for a large and diverse dataset, especially for severe and PDR cases, emphasizes the resource-intensive nature of our proposed methodology. These factors collectively contribute to the difficulty in DR stage grading. A visual representation of the model's predictions for the test dataset, including both correct classifications and misclassifications, is depicted in Fig.~\ref{fig:fig13}, providing further insight into the challenges faced by the model in accurately classifying DR severity levels.

To address the weaknesses of our model, several potential solutions can be explored. One approach could be integrating additional imaging modalities, such as optical coherence tomography (OCT) scans, which offer more detailed information about retinal structures. The combination of fundus images with OCT scans could enhance the model's ability to accurately classify severe and PDR cases by providing supplementary insights into the retinal architecture. Another potential solution could be to collect more data for these categories, particularly in cases with varying severity, to improve the robustness of the model. Additionally, incorporating a more extensive range of clinical features, such as the duration of diabetes, blood pressure measurements, and other relevant medical history indicators could serve as valuable additional inputs. Furthermore, leveraging advanced and specialized preprocessing techniques may yield further improvements in the model's performance.

Despite the mentioned limitations, our proposed model still outperformed existing methods in the literature. To ensure a fair evaluation, we compared our proposed model with existing methods that evaluated their works using a randomly selected subset of the APTOS 2019 dataset in Table \ref{tbl5}.  

In binary classification, our proposed dual-branch model achieved an outstanding accuracy of 98.50\%, positioning it as one of the top-performing approaches. The proposed method exhibited a remarkable sensitivity of 99.46\%, highlighting its exceptional ability to accurately identify positive cases. While the precision of Islam et al. \cite{Islam2022} for binary classification slightly exceeded our model, our proposed approach demonstrated strong overall performance, underscoring its suitability for diabetic retinopathy detection. For multi-class classification, our proposed model achieved an impressive accuracy of 89.60\%. This performance surpasses that of previous methods, even those based on transfer learning. These results highlight the effectiveness and superiority of our approach, which incorporates several key elements.

Firstly, the selection of feature extractors plays a vital role in the performance of our model. By carefully choosing and integrating powerful feature extractors, namely EfficientNetB0 and ResNet50, we have been able to capture and represent the discriminative features present in the input images more effectively. 

Moreover, our utilization of image augmentation techniques has contributed significantly to the success of our model. By applying various transformations, we have effectively augmented the dataset, thereby enhancing the model's ability to generalize and make accurate predictions on unseen data. This augmentation process has likely contributed to the improved accuracy achieved by our model in multi-class classification.

Additionally, our approach of selectively merging public datasets for training our model has proven advantageous. By integrating multiple datasets containing diverse samples, we have increased the diversity and richness of the training data. This augmented dataset provides our model with a more comprehensive understanding of different DR severity levels, leading to improved classification accuracy.

\begin{table*}[b]
 \caption{Comparison of state-of-the-art methods with our proposed model for binary and multi-class classification with APTOS 2019.}\label{tbl5}
 \begin{tabularx}{\textwidth}{Xcccc}
 \toprule
 Reference number & Method & Accuracy (\%) & Precision (\%) & Sensitivity (\%) \\
 \midrule
    \textbf{Binary classification} & & & & \\
    \cite{Bodapati2020} & Blended features + DNN & 96.10 & - & - \\
    \cite{Kumar2021} & Hybrid model (VGG16 + Capsule network) & 97.05 & - & - \\
    \cite{Islam2022} & Supervised contrastive learning & 98.36 & \textbf{98.36} & 98.37 \\
    Proposed model & Deep dual-branch model (transfer learning) & \textbf{98.50} & 97.61 & \textbf{99.46} \\
    \textbf{Multi-class classification} &  &  & &\\
    \cite{Dondeti2020} & NASNet + $\nu$-SVM & 77.90 & 76.00 & 77.00 \\
    \cite{Kassani2019} & Modified Xception & 83.09 & -  & 88.24 \\
    ResNet50 \cite{Kassani2019} & Original ResNet50 & 74.64 & -  & 56.52 \\
    Xception \cite{Kassani2019} & Original Xception & 79.59 & -  & 82.35 \\
    InceptionV3 \cite{Kassani2019} & Original InceptionV3 & 78.72 & -  & 63.64 \\
    MobileNet \cite{Kassani2019} & Original MobileNet & 79.01 & -  & 84.62 \\
    \cite{Gangwar2021} & Hybrid model with Inception-ResNet-v2 & 82.18 & - & - \\
    \cite{Dekhil2019} & Custom CNN & 77.00 & - & - \\
    \cite{Bodapati2020} & Blended features + DNN & 80.96 & - & - \\
    \cite{Kumar2021} & Hybrid model (VGG16 + Capsule network) & 75.50 & - & - \\
    \cite{Islam2022} & Supervised contrastive learning & 84.36 & 70.51 & 73.84 \\
    Proposed model & Deep dual-branch model (transfer learning) & \textbf{89.60} & \textbf{89.23} & \textbf{89.60} \\
\bottomrule 
\end{tabularx}
\end{table*}

\section{Conclusion and future work}
In this paper, we presented a novel dual-branch transfer learning approach for the detection and grading of various stages of diabetic retinopathy using a single retinal fundus photograph. Our model utilizes two state-of-the-art CNN architectures, EfficientNetB0 and ResNet50, which have been fine-tuned through transfer learning techniques on a large multi-center dataset. A distinctive feature of our work lies in curating our training dataset. We combined data from three public datasets and performed a meticulous merging and augmentation process, selectively incorporating categories based on insights gained from extensive experiments. This careful dataset curation serves as a key contribution, setting our model apart from other existing methods in the field. The proposed approach demonstrates superior performance compared to existing methods in the literature, as validated by a diverse set of evaluation metrics. We also employed the CCE loss function to enhance the robustness of our model against data imbalances, and its effectiveness is evident in our results. 

Nevertheless, as explained in Section {\ref{sec:discussion}}, there exists potential for enhancement through future research. Future research directions may involve investigating advanced preprocessing techniques, further hyperparameter tuning, exploring other state-of-the-art models as feature extractors, and examining the use of additional modalities to enhance the accuracy of DR diagnosis. Our results suggest that the proposed model can serve as a reliable tool for clinical decision-making and can assist ophthalmologists in the diagnosis and grading of DR.  Ultimately, we believe that our proposed approach can be extended to other medical imaging tasks beyond diabetic retinopathy. This study makes a valuable contribution to the increasing body of literature on transfer learning-based methods for medical image analysis, and we hope that our work inspires further exploration in this area.


\printcredits

\section*{Funding}
The research conducted for this paper did not receive any financial support or resources from any external sources.

\section*{Conflict of Interest}
The authors declare no conflict of interest.

\section*{Declaration of Generative AI and AI-assisted technologies in the writing process}
During the preparation of this work, the authors used ChatGPT\footnote{Official website: \href{https://chat.openai.com/}{https://chat.openai.com/}} for text proofreading and writing improvement purposes. After using this tool, the authors reviewed and edited the content as needed and take full responsibility for the content, data analysis, and interpretations of the publication.



\bibliography{refs}

\end{document}